\documentstyle[prl,aps,epsfig]{revtex}
\begin{document}
\draft
\twocolumn[\hsize\textwidth\columnwidth\hsize\csname@twocolumnfalse\endcsname
\title{Hydrodynamics of thermal granular convection}
\author{Xiaoyi He$^{1}$, Baruch Meerson$^{2}$ and Gary Doolen$^{1}$}
\address{$^1$Los Alamos National Laboratory, Los Alamos, NM 87545}
\address{$^2$Racah Institute of Physics, Hebrew University of
Jerusalem, Jerusalem 91904, Israel}
\date{\today}
\maketitle
\begin{abstract}
A hydrodynamic theory is formulated for buoyancy-driven (``thermal") granular convection, recently
predicted in molecular dynamic simulations and observed in experiment. The limit of a dilute flow
is considered. The problem is fully described by three scaled parameters. The convection occurs via
a supercritical bifurcation, the inelasticity of the collisions being the control parameter.  The
theory is expected to be valid for small Knudsen numbers and nearly elastic grain collisions.
\end{abstract}
 \vskip1pc] \narrowtext

As we know from experience, hot fluid rises. Is the same statement true for \textit{granular}
fluid, where the role of temperature is played by fluctuations of the grain velocities? There is
strong recent evidence that the answer to this question is positive. Buoyancy-driven ``thermal"
granular convection was first observed in molecular dynamic (MD) simulations of granular gas in
two dimensions \cite{Ramirez}, with no shear or time-dependence introduced by the system
boundaries. It was found that thermal granular convection appears via a supercritical bifurcation,
with inelastic collision losses being the control parameter \cite{Ramirez}. Strong evidence for
thermal granular convection was recently obtained in experiment with a highly fluidized 3D granular
flow \cite{Wildman} (see also an earlier work \cite{Bizon}). In these two systems
\cite{Ramirez,Wildman} the convection is driven by a negative vertical granular temperature
gradient \cite{temperature} which makes this convection similar to the classical
Rayleigh-B\`{e}nard convection \cite{Chandrasekhar} and its analogs in compressible fluid
\cite{Spiegel,GS,G,CU}. In the Rayleigh-B\`{e}nard problem a negative temperature gradient is
imposed externally. In a granular flow driven from below, it develops spontaneously because of the
inelasticity of particle collisions \cite{walls}.

The phenomenon of ``thermal" convection in granular fluids is fascinating, as it gives one more
example of similarities/differences between the granular and ``classical" fluids \cite{Kadanoff}.
Though basic properties of thermal granular convection were investigated in the MD simulations
\cite{Ramirez}, no theory exists yet. The objective of the present work is to formulate such a
theory. We will work in the regime where the ``standard" granular hydrodynamic equations in 2D
\cite{Jenkins}, systematically derivable from kinetic equations \cite{Goldhirsch},  are expected
to be accurate. As it has become clear by now \cite{Goldhirsch},  this requires (in addition to
the strong inequality $K\ll 1$, see below, and sufficiently low density) that particle collisions
be nearly elastic: $q \ll 1$, where $q=(1-r)/2$ is the inelasticity coefficient and $r$ is the
normal restitution coefficient. The nearly elastic limit is motivated by the MD simulations
\cite{Ramirez} where convection was observed already at very small inelasticities: $4 \times
10^{-4} \le q \le 2 \times 10^{-2}$. As in the MD simulations \cite{Ramirez}, we will assume a
velocity-independent restitution coefficient. We will limit ourselves to dilute flow, $n \ll n_c$,
where $n$ is the number density of the grains and $n_c$ is the close-packing density. As thermal
granular convection does not necessarily involve clustering, the latter assumption is not too
restricting.

Here is an outline of the rest of this Communication. We will see that the hydrodynamic problem of
thermal granular convection is fully determined by \textit{three} scaled parameters: the Froude
number $F$, Knudsen number $K \ll 1$ and inelasticity coefficient $q \ll 1$, and by the aspect
ratio of the system. The translationally symmetric static steady state of the system plays the role
of the ``simple conducting state" of the Rayleigh-B\`{e}nard problem. By employing the Lagrangian
mass coordinate, we find this steady state analytically.  Then, by solving the granular
hydrodynamic equations  in a square box by a Lattice-Boltzmann method, we observe a supercritical
bifurcation at a critical value of $q=q_c$ and steady convection at $q>q_c$, in qualitative
agreement with MD simulations \cite{Ramirez} and experiment \cite{Wildman}. We then investigate
the dependence of the convection threshold $q_c$ on $K$ and $F$. Our results open the way to a
systematic investigation of thermal granular convection.

Let $N \gg 1$ identical smooth hard disks with diameter $d$ and mass $m$ move without friction and
inelastically collide inside a two-dimensional box with lateral dimension $L_x$ and height $L_y$.
The aspect ratio of the system $\Delta=L_x/L_y$. The gravity acceleration $g$ is in the negative
$y$ direction. The particles are driven by a base that is kept at temperature $T_0$. For
simplicity, the three other walls are assumed elastic. The hydrodynamic description deals with
coarse-grained fields: the number density of grains $n (\textbf{r},t)$, granular temperature $T
(\textbf{r},t)$ and mean velocity of grains $\textbf{v} (\textbf{r},t)$. The governing equations
can be written, in the dilute limit, in the following scaled form:
\begin{equation}
\frac{d n}{d t} + n\, {\bf \nabla} \cdot {\bf v} = 0\,, \label{cont}
\end{equation}
\begin{equation}
n \, d{\bf v}/dt = {\bf \nabla} \cdot {\bf P} - F \, n\, {\bf e_y}\, \label{momentum}
\end{equation}
$n\, dT/dt + nT\, {\bf \nabla} \cdot {\bf v} =$
\begin{equation}
K\left[{\bf \nabla} \cdot (T^{1/2} {\bf \nabla} T) -R\, n^{2}\, T^{3/2}\right]\,. \label{heat}
\end{equation}
Here  ${\bf e_y}$ is the unit vector along $y$, $d/dt$ is the total derivative, ${\bf P} = - nT
\,{\bf I} + \frac{1}{2}\, K\, T^{1/2}\, \hat{{\bf D}}$ is the stress tensor, ${\bf D}
=(1/2)\left({\bf \nabla v} + ({\bf \nabla v})^{T}\right)$ is the rate of deformation tensor,
$\hat{{\bf D}}={\bf D}-(1/2)\, tr\, ({\bf D}\,)\, {\bf I}$ is the deviatoric part of ${\bf D}$ and
${\bf I}$ is the identity tensor. In the dilute limit the bulk viscosity can be neglected compared
to the shear viscosity \cite{Jenkins}. In addition, we neglected the small viscous heating term in
the heat balance equation (\ref{heat}). The three scaled parameters entering Eqs. (\ref{momentum})
and (\ref{heat}) are the Froude number $F=m g L_y/T_0$, the Knudsen number $K=2
\pi^{-1/2}\,\left(d\,L_y\, \langle n\rangle\right)^{-1}$, and the collision losses parameter
$R=8q\,K^{-2}$. $R$ will be used through the rest of this Communication instead of the
inelasticity coefficient $q$. The Knudsen number $K$ is of order of the ratio of the mean free path
of the grains to the system height. For hydrodynamics to be valid we should demand $K \ll 1$. The
units of distance, time, velocity, density and temperature in the scaled equations are $L_y$,
$L_y/T_0^{1/2}$, $T_0^{1/2}$, $\langle n \rangle$ and $T_0$, respectively. Finally, $\langle n
\rangle =N/(L_x L_y)$ is the average number density of the grains.

The physical meaning of the scaled numbers $F$, $K$ and $R$ is clear. The Froude number $F$
determines the relative role of the maximum potential energy of grains in the gravity field and
their maximum fluctuation energy supplied by the driving base. The Knudsen number $K$ determines
the efficiency of the momentum and energy transport in the system. In the hard-sphere fluid we are
working with, the kinematic viscosity and thermal diffusivity are equal to each other, so the
Prandtl number is equal to 1. The inelastic heat loss number $R$ determines the relative role of
the inelastic heat losses and heat conduction.

The boundary conditions are $T(x,y=0,t) = 1$ with a zero heat flux at the rest of the boundaries.
Also, we demand zero normal components of the velocity and slip (no-stress) conditions at all
boundaries. The total number of particles is conserved:
\begin{equation}
\frac{1}{\Delta} \int_0^{\Delta} dx\,\int_0^1 dy\, n(x,y,t) = 1\,. \label{conservation}
\end{equation}
Therefore, in the hydrodynamic formulation, the problem is characterized by $F,K$ and $R$ and the
aspect ratio $\Delta$, instead of the full set of 8 parameters $m,d,q,L_x,L_y, g, T_0$ and $N$.

Translationally symmetric static steady states are described by the one-dimensional equations
considered in many works (see, e.g. Ref. \onlinecite{Eggers}):
\begin{equation}
(n_s\,T_s)^{\prime} + F\, n_s = 0 \label{static1}
\end{equation}
and
\begin{equation} (T_s^{1/2} T_s^{\prime})^{\prime} - R\, n_s^{2}
T_s^{3/2} = 0 \label{static2}
\end{equation}
(primes denote $y$-derivatives). In our problem these equations are complemented by the boundary
conditions $T_s(y=0)=1$ and $T_s^{\prime} (y=1) = 0$ and normalization condition $\int_0^1 dy\,
n_s (y)=1$ . A static state is characterized by the scaled numbers $F$ and $R$. Equations
(\ref{static1}) and (\ref{static2}) can be solved by going over to the Lagrangian mass coordinate
$\mu (y)=\int_0^y n_s(y^{\prime})\,dy^{\prime}$ \cite{Zeldovich}. In view of Eq.
(\ref{conservation}), the Lagrangian mass coordinate $\mu$ changes between $0$ and $1$. First, we
solve Eq. (\ref{static1}) for $n_s (\mu)$:
\begin{equation}
n_s(\mu)=\frac{p_0-F\,\mu}{T_s (\mu)}\,, \label{density}
\end{equation}
where $p_0$ is the (as yet unknown) pressure at the thermal base $\mu=0$. Substituting this
relation into Eq. (\ref{static2}), we obtain a \textit{linear} equation for $Y(\mu)\equiv
T_s^{1/2}(\mu)$:
\begin{equation}
(\lambda-\mu)\, Y^{\prime\prime} - Y^{\prime} - (R/2)\,(\lambda-\mu)\, Y =0\,, \label{lineq}
\end{equation}
where $\lambda=p_0/F$ and the primes now stand for $\mu$-derivatives. The general solution of Eq.
(\ref{lineq}) is a linear combination of the Bessel functions $I_0[\sqrt{R/2} (\lambda-\mu)]$ and
$K_0[\sqrt{R/2} (\lambda-\mu)]$. The two arbitrary constants are found from the boundary
conditions $Y(\mu=0)=1$ and $Y^{\prime}(\mu=1)=0$. Now we employ Eq. (\ref{density}) for
$n_s(\mu)$ and determine the Eulerian coordinate $y=y(\mu)$ by calculating $y=\int_0^{\mu}
d\mu'/n_s(\mu')$. Demanding that $y(\mu=1)=1$, we find $\lambda$ (and, therefore, $p_0$) which
completes the solution. An example of static temperature and density profiles is shown in Fig.
\ref{fig1}. We found that, at $F=0.1$ and $R \lesssim 0.7$, the temperature difference between the
lower and upper plates in the static solution agrees very well with the MD simulations results
\cite{Ramirez}. The negative temperature gradient is clearly seen in Fig. 1. At sufficiently large
$R$, a denser and heavier gas is located on top of the underdense gas. This destabilizing factor
drives convection. The stabilizing factors are granular viscosity and heat conduction.

We investigated convective (in)stability of the static state by solving the time-dependent
hydrodynamic equations (\ref{cont})-(\ref{heat}) numerically. A lattice-Boltzmann scheme,
previously used to study the classical Rayleigh-B\`{e}nard convection \cite{He98}, was employed.
The scheme give accurate results for moderate density variations which was the case in the
parameter range of this study. Like in the MD simulations \cite{Ramirez}, we considered a square
box: $\Delta=1$. The initial conditions were the following: a uniform (and equal to 1) temperature,
zero velocity and density equal to $1$ plus a small sinusoidal perturbation. We fixed $F$ and $K$
and varied $R$. The presence (absence) of convection in the box was measured by computing (after
transients die out) the velocity circulation $C=\oint {\bf v} \cdot {\bf dl}$ along the edges of
the box. In all cases a zero circulation is observed at sufficiently small $R$, and the flow
approaches a static steady state. We checked that the density and temperature profiles of the
steady state, obtained in the lattice-Boltzmann simulations, agree within 1.5 \% with the analytic
solutions of Eqs. (\ref{static1}) and (\ref{static2}). Convection always develops, via a
supercritical bifurcation, when $R$ exceeds a critical value $R_c (F,K)$. Figure \ref{fig3} shows
steady convection that appears in this system after transients decay. Figure \ref{fig4} shows the
bifurcation diagram. The same type of bifurcation (supercritical bifurcation) was observed in the
MD simulations \cite{Ramirez}.

\begin{figure}[h]
\centerline{ \epsfig{file=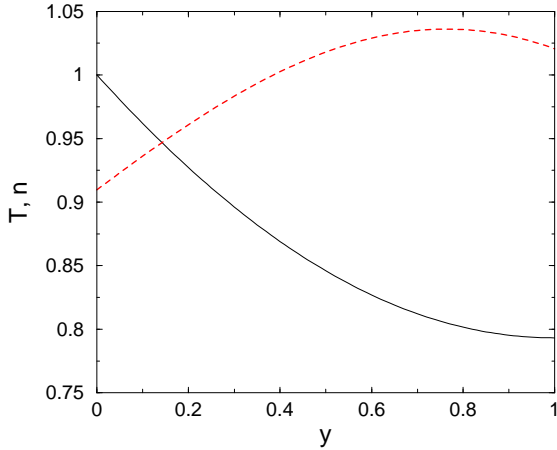, width=2.9in, clip= }} \caption{One-dimensional static
temperature (solid line) and density (dashed line) profiles for $F=0.1$ and $R=0.5$.} \label{fig1}
\end{figure}

\vspace{0.4cm}
\begin{figure}[h]
\centerline{ \epsfig{file=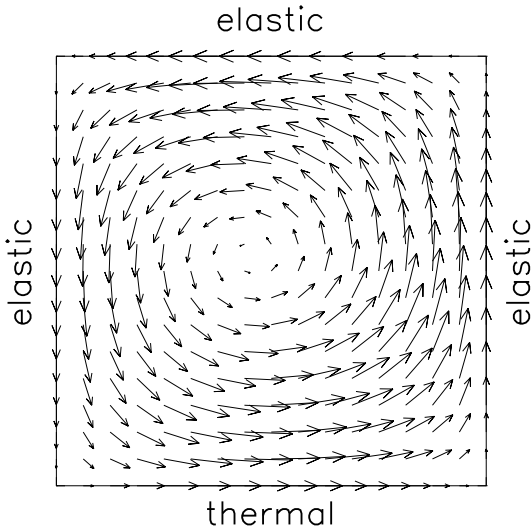, width=3.5in}}\vspace{0.3cm} \caption{Steady-state hydrodynamic
velocity field for $F=0.1$, $K=0.02$ and $R=3.6$.} \label{fig3}
\end{figure}

We determined the convection onsets and bifurcation diagrams for two values of the Froude number:
$F=0.05$ and  $0.1$, varying the Knudsen number $K$ between $0.01$ and $0.06$. The results of this
series of simulations, depicted in Fig. \ref{fig4a}, clearly  show that the viscosity and heat
conduction (both of which scale like $K$) are stabilizing factors. Also, it can be seen that
stronger gravity promotes convection as expected.

\begin{figure}
\centerline{\epsfig{file=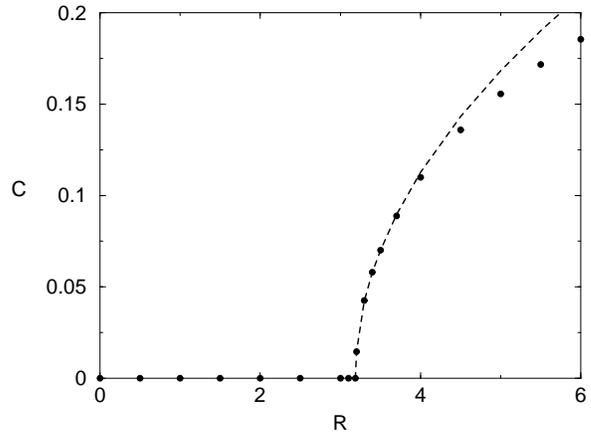, width=3in}} \caption{Velocity circulation $C$ along the edge of
the box vs. $R$ measured in hydrodynamic simulations (points), and the curve
$0.125\,(R-3.186)^{1/2}$ (dashed line). In this example $F=0.1$ and $K=0.05$.} \label{fig4}
\end{figure}

\begin{figure}
\centerline{ \epsfig{file=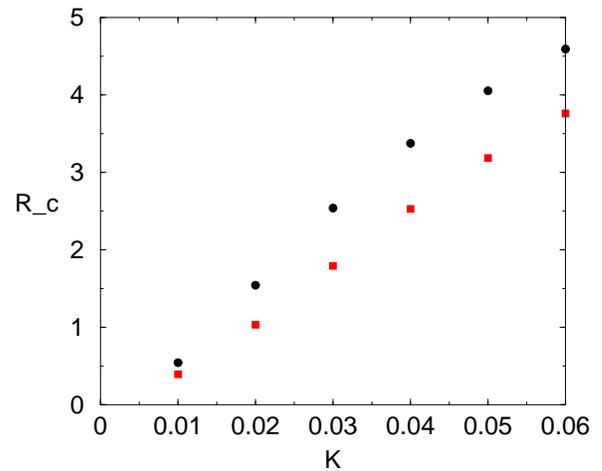, width=3in, clip= }} \caption{The critical value $R_c$ for the
convection onset vs. the Knudsen number $K$ for $F=0.05$ (circles) and $0.1$ (squares).}
\label{fig4a}
\end{figure}

Transient motions in the system were investigated by monitoring the maximum value of the
hydrodynamic velocity in the system as a function of time. After initial transients decay, the
dynamics depends on whether one is in the subcritical ($R < R_c$) or supercritical ($R > R_c$)
range. We found that, in the supercritical range, the maximum velocity first increases
exponentially in time (no ``overstability"), and then approaches, in an oscillatory way, a constant
value corresponding to a steady convection.  We used the growth rates to extrapolate to the
critical values $R_c$ for the convection onset. We also found that the frequency of the decaying
oscillations around the steady convection vanishes at the convection onset.

The next theoretical step should be the linear stability analysis of Eqs.
(\ref{cont})-(\ref{heat}) around the static solutions, and a detailed investigation of the
hydrodynamic modes of the system. In the spirit of pattern formation theory \cite{Cross}, one
should also study convection in a strip infinite in the lateral direction, by varying the lateral
wave number of the perturbation. This analysis is presently under way.% Similar to compressible
%atmospheres in astrophysics \cite{Moore}, a hydrodynamic system of this order can have up to five
%collective modes. At sufficiently small $K$ and $R$ and at moderate $F$, two pairs of modes
%represent (slightly damped) high-frequency oscillatory modes: the gravoacoustic modes modified by
%heat conduction and inelastic heat losses. An additional mode is a low-frequency thermal mode. Our
%numerical results imply that this mode is damped at $R<R_c$. At $R>R_c$ it becomes unstable, and
%instability is aperiodic on the onset. The absence of overstability apparently results from the
%big difference between the frequencies of the fast and slow modes of the system, similarly to the
%``classical" compressible convection \cite{Spiegel,GS,G}.

Observing thermal granular convection in experiment requires several conditions, some of which can
be stringent. A detailed discussion of these conditions is beyond the scope of this Communication.
However, there is one crucial issue that has to be discussed. A standard method of fluidization of
granular materials (used, in particular, in experiment \cite{Wildman}) is vibration of the bottom
plate. There are two important \textit{necessary} conditions for observing thermal granular
convection in vibrofluidized granular beds. Firstly, the frequency of vibration of the bottom
plate should be much higher than any relevant \textit{macroscopic} frequency of the granulate (like
the frequency of the bed oscillations or inverse sound travel time). Secondly, the vibration
amplitude should be less than the mean free path of the granulate near the bottom wall. These
conditions guarantee that there is no direct coupling between the bottom plate vibration and
collective granular motion. Additional conditions are those of convection instability. Our theory
gives such a condition: $R>R_c\, (K,F)$. However, we obtained this condition (a) in the dilute
limit $n \ll n_c$, and (b) for a "thermal", rather than vibrating, bottom plate. Limitation (a) can
be severe unless the experiment is done in a 2D geometry (spherical particles rolling on a
slightly inclined smooth surface and driven by a vibrating wall \cite{Kudrolli}). To what extent
is limitation (b) severe? A full quantitative answer to this question requires solving a similar
hydrodynamic problem, but with a different boundary condition \cite{Eggers,Kumaran} that mimics
the vibrating wall more directly. Based on an analogy with other driven granular systems
\cite{LMS}, we expect that the results obtained for the two boundary conditions will not differ
too much from each other, at least qualitatively.

Finally, it is possible that the concept of ``thermal" convection can be applicable to some
granular systems driven by shear. A recent example is the longitudinal vortices observed in
experiment on rapid granular flow in a chute \cite{Forterre}.

To summarize, granular hydrodynamics provide a proper language for the problem of thermal granular
convection and open the way to a systematic investigation of this and related phenomena.

We acknowledge useful discussions with John M. Finn, Jerry P. Gollub, Evgeniy Khain, Rosa
Ram\`{\i}rez and Victor Steinberg. B.M. is very grateful to the Center for Nonlinear Studies of
Los Alamos National Laboratory, where this work started, for hospitality and support. The work was
supported in part by the Israel Science Foundation administered by the Israel Academy of Sciences
and Humanities.


\begin{references}
\bibitem{temperature} The (local) granular temperature is defined
as the mean fluctuation of the (local) kinetic energy of the grains: $\langle (m/2) \,({\bf
v}_p-{\bf v})^2 \rangle$, where ${\bf v}$ is the mean velocity.
\end{references}
\end{document}